\documentclass[11pt]{article}

\usepackage{amssymb}
\usepackage{amsmath}
\usepackage{graphicx}
\usepackage{epsfig}
\usepackage{amsfonts}
\usepackage{color}
\usepackage{float}
\usepackage{latexsym}
\usepackage{mathrsfs}
\usepackage{feynmf}
\usepackage{balance}

\def\be{\begin{equation}} \def\ee{\end{equation}}
\def\ba{\begin{eqnarray}} \def\ea{\end{eqnarray}} \def\part{\partial}

   \def\b1{{\bf 1}}

\begin{document}

\begin{titlepage}
\title{
\begin{flushright}\begin{small}
\end{small}\end{flushright}\vspace{2cm}
Soliton-potential Interaction in the Nonlinear Klein-Gordon Model}

\author{Danial Saadatmand\thanks{Email: Da$\_$se.saadatmand.643@stu-mail.um.ac.ir} $\,$ and
Kurosh Javidan\thanks{Email: Javidan@um.ac.ir} \\ \\
\small{Department of physics, Ferdowsi university of Mashhad}\\
\small{91775-1436  Mashhad Iran} }

\maketitle

\abstract{Interaction of solitons with external potentials in the nonlinear Klein-Gordon field theory is investigated using an improved model. The presented model is constructed with a better approximation for adding the potential to the Lagrangian through the metric of background space-time. The results of the model are compared with another model and the differences are discussed.}
\end{titlepage}
\setcounter{page}{2}
\section{Introduction}
\quad\, Nonlinear evolution equations in mathematical physics are largely studied because of their important roles in most of the branches of science.The Klein-Gordon (KG) equation  is a relativistic version of the Schrodinger's equation which describes field equation for scalar particles (spin-0). The KG equation has been the most frequently studied equation for describing the particle dynamics in quantum field theory. This equation with various types of potentials has appeared in field theories which is called nonlinear Klein-Gordon (NKG) model. Pion form factor \cite{i1}, Hadronic atoms \cite{i2}, Josephson junction \cite{i3}, Superfluid current disruption \cite{i4}, Shallow water \cite{i5} and Hawking radiation from a black hole \cite{i6} are some examples of the NKG applications.

In recent years several solitonic solutions for the NKG equations have been proposed using different methods. These solutions have appeared in homogeneous and well behaved medium while in real world the medium of propagation contains disorders and impurities which add local space-dependent potentials to the problem. The behaviour of the NKG solitons during the interaction with these local potentials and their stability after the interaction are important questions. Therefore, investigating these situations is very interesting. It is a very important subject because of its applications and also due to mathematical point of view. Motivated by this situation, we have studied the interaction of the NKG solitons with defects using different methods and the results are presented in this paper. Therefore a brief description of the NKG model and its solitons has been introduced in section 2. The motivation of the present work and some applications of the model in the presence of local potentials will be proposed in this section, too. Methods of adding the potential to the soliton equation of motion will be described in section 3. Interaction of solitons with the potential walls and potential wells will be investigated in section 4. Some conclusions and remarks will be presented in the final section.   

%====================================================================

\section{Nonlinear Klein-Gordon Equation} \setcounter{equation}{0}
\quad\, Different types of nonlinear Klein-Gordon equation have been proposed. They are different in their nonlinear terms which arise from application manners. Some of them are as follows: 

\begin{eqnarray} \label {KG eqs}
&& \frac{\partial^{2}\phi}{\partial t^{2}}-a^{2}\frac{\partial^{2}\phi}{\partial x^{2}}+\alpha \phi-\beta \phi^{3} =0 \\ 
&& \frac{\partial^{2}\phi}{\partial t^{2}}-a^{2}\frac{\partial^{2}\phi}{\partial x^{2}}+\alpha \phi-\beta \phi^{3}+\gamma \phi^{5}=0,  \\ \nonumber 
\end{eqnarray}
in which  $\alpha$ and $\beta$ are arbitrary constants and $\gamma=\frac{3\beta^{2}}{16 \alpha}$. Several localized solutions such as solitons, compactons and solitary waves have been found for these equations using different methods \cite {i7,i8,i9,i10}. In this paper we will focus on the soliton solutions of the equation (2.2). The results are valid for the other equations, too.

Wazwaz has found several localized solutions for the equation (2.2) using "Tanh" method. One soliton solution for this equation is  \cite{i7}
\begin{equation}\label{kg 1s}
\phi(x,t)=\sqrt{\frac{2\alpha}{\beta}\left[1\pm tanh\left(\sqrt{\frac{\alpha}{a^{2}-u^{2}}}(x-x_{0}-ut) \right)\right]},   
\end{equation}
where $x_{0}$ is the soliton initial position and $ u $ is its velocity. The signs "+" and "-" in the equation (\ref{kg 1s}) denote to the kink and antikink solutions respectively. Figure 1 presents kink and antikink solutions with $u=0.5$, $a=1$, $ x_{0}=0$ and $\alpha=\beta=1$ at $t=0$.

\begin{figure}[htbp]
\begin{center}
\leavevmode \epsfxsize=9cm \epsfbox {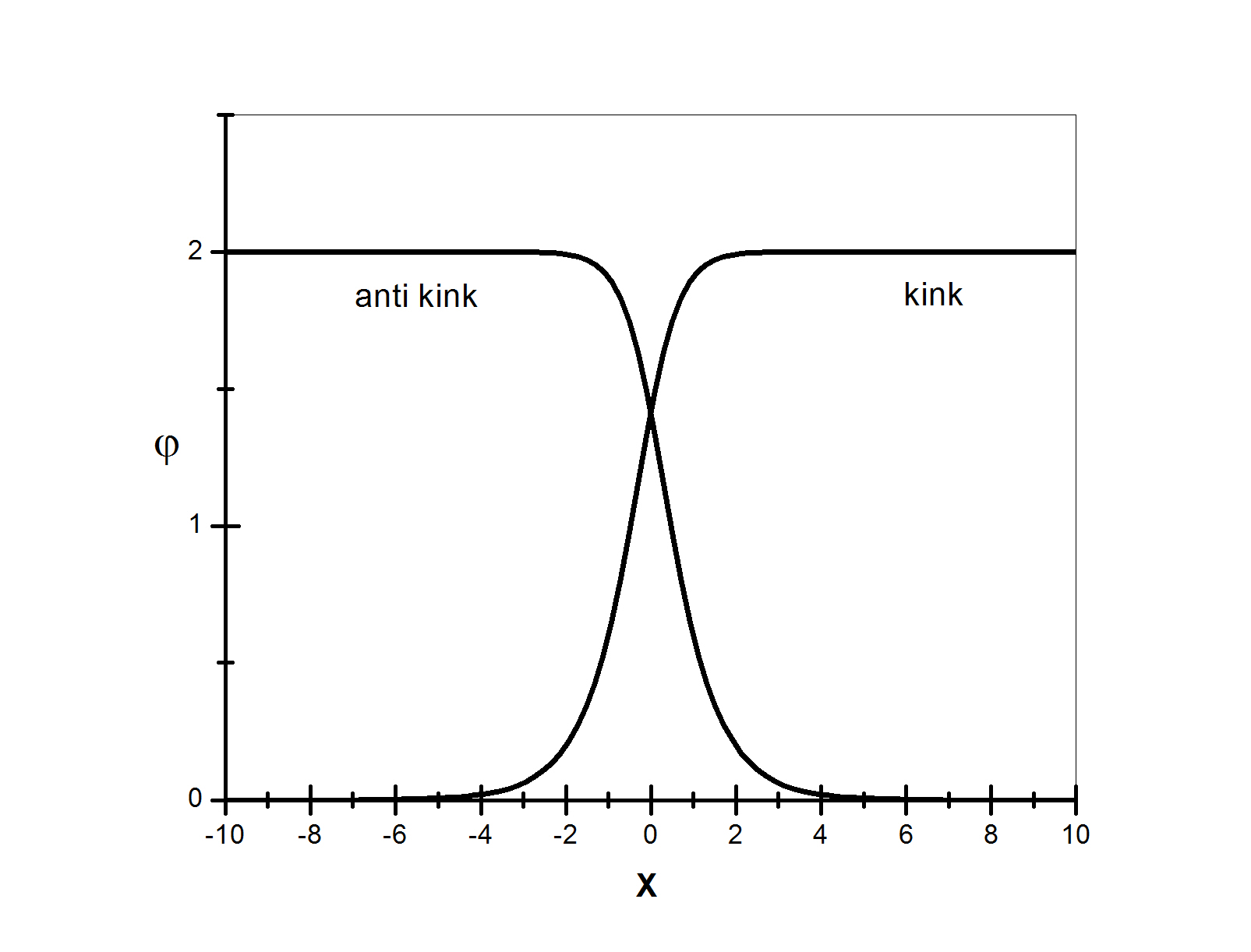}
\end {center}
\caption{Soliton solution of the NKG described by equation (\ref{kg 1s}).}
\label{fig1}
\end{figure}

As mentioned before, the Klein-Gordon type field theories have been applied to many investigations. The quantum aspects of the black hole have been studied considering a time-dependent classical black hole solution induced by the Klein-Gordon soliton \cite{i6}.The Hawking radiation and the Bondi reflected energy of the back reaction of the metric were calculated by adding a lump of energy through the Klein-Gordon field as
\begin{equation}\label{black}
U\left (\phi\right )=\frac{\beta}{4}\left(\phi^{2}-\frac{\mu^{2}}{\beta}\right)^{2}. 
\end{equation}
The parameters $\beta$ and $\mu$ are constant if soliton is sufficiently narrow. In this situation the effects of the gravity on the soliton characters are neglected. For broader energy lumps, we cannot take constant parameters for the soliton and its characters which are spatial functions\cite{i6a, i6b}. Therefore, a better situation can be investigated if we can consider a potential for the Klein-Gordon part of the theory with space dependent parameters like $\beta(x)$ and $\mu (x)$. 

The behaviour of transition regions in one-dimensional multistable continuous media is described by the nonlinear Klein-Gordon equation with dissipation (NKGD) \cite{i6c}
\begin{equation}\label{cont}
 \mu \phi_{tt} + \gamma \phi_{t}-D\phi_{xx} + P(\phi)=0,   
\end{equation}
 where $P(\phi)$ is a function of the field $\phi$. 
 Generally, the parameters of the model are some functions of the medium characters like temperature, viscosity and so on; however, such a system has not been studied  with constant parameters yet.  It is clear that we need a suitable model containing variable (at least slowly varying) parameters.
 
 Much work has been devoted to the dynamics of the one dimensional (1D) chain of coupled double-well oscillators and its applications in condensed matter \cite{i6c1,i6c2,i6c3}. This model as well as the Frenkel-Kontorova chain became a very convenient and popular tool for the theoretical studies of dynamical and statistical properties of a number of materials with strong quasi-1D anisotropy \cite{i6c4}. One of the interesting systems of this class is a chain of hydrogen bonds where the proton transfers in adjacent hydrogen bonds are correlated because of the cooperative nature of hydrogen bonding. The proton dynamics in an ice crystal can be described using the discrete nonlinear Klein-Gordon (NKG) equation\cite{i6d, i6e}
\begin{equation}\label{ice}
\phi_{tt}  - c_{0}^{2} \phi_{xx} + \omega_{0}^{2} V(\phi)=0.   
\end{equation} 

It is clear that a more realistic model contains disorders and dislocations in the hydrogen chain, which can be modeled using space dependent parameters in the NKG field theory.

%=========================================================================

\section{Two Models for "the NKG Soliton-potential" System }
\quad\, \textbf{Model 1}: Consider a scalar field with the Lagrangian 
\begin{equation}\label{lag}
{\cal L}=\frac{1}{2}\partial_ {\mu}\phi\partial^{\mu}\phi-U\left (\phi\right )   
\end{equation}
and the following potential: 
\begin{equation}\label{uphi}
U\left (\phi\right )=\lambda (x)\left(\frac{1}{2}\alpha \phi^2-\frac{1}{4}\beta \phi^{4} +\frac{1}{6}\gamma \phi^{6}\right).   
\end{equation}
The equation of motion for the field becomes
\begin{equation}\label{eq}
\partial_ {\mu}\partial^{\mu}\phi+\lambda(x)\left(\alpha \phi-\beta\phi^3+\gamma \phi^5\right )=0.  
\end{equation}
The effects of the external potential can be added to the equation of motion by using a suitable definition for $\lambda(x)$, like $\lambda(x)=1+v(x)$ \cite{i11,i12,i13}. For a constant value of parameter ($\lambda(x)=1$) equation (\ref{eq}) reduces to (2.2) and therefore (\ref{kg 1s}) is its exact solution. Thus (\ref{kg 1s}) is used as an initial condition for solving (\ref{eq}) with a space dependent $\lambda(x)$ when the potential $v(x)$ is small. 

Hamiltomian density of the Lagrangian (\ref{lag}) is
\begin{equation}\label{h1}
{\cal H}_{1}=\frac{1}{2}\dot{\phi}^2+\frac{1}{2}\acute{\phi}^2+\lambda(x)\left(\frac{1}{2}\alpha \phi^2-\frac{1}{4}\beta \phi^{4} +\frac{1}{6}\gamma \phi^{6}\right).
\end{equation}

\textbf{Model 2}: The potential also can be added to the Lagrangian of the system, through the metric of background space-time. So the metric carries the medium characters. The general form of the action in an arbitrary metric is
\begin{equation}\label{action metric}
I=\int{{\cal L}(\phi , \partial_{\mu}\phi)\sqrt{-g}d^{n}x dt },
\end{equation}
where "g" is the determinant of the metric $g^{\mu\nu}(x)$. A suitable metric in the presence of a smooth and slowly varying weak potential $v(x)$ is \cite{i14,i15}
\begin{equation}\label{new metric}
g^{\mu \nu}(x)\cong\left(
\begin{array}{clrr} 1+v(x) & 0 \\ 0 & -\frac{1}{1+v(x)}
\end{array}\right).
\end{equation}  
The equation of motion for the field $\phi$ which is described by the Lagrangian (\ref{lag}) in the action (\ref{action metric}) is \cite{i15,i16}
\begin{equation}\label{Eq motion}
\frac {1}{\sqrt{-g}}\left (\sqrt{-g}\partial_{\mu}\partial^{\mu}\phi+\partial_{\mu}\phi\partial^{\mu}\sqrt{-g}\right )+\frac {\partial U(\phi)}{\partial \phi}=0.
\end{equation} 
This equation of motion in the background space-time (\ref{new metric}) becomes \cite{i17}
\begin{equation}\label{new Equation motion}
\left ( 1+v(x)\right )\frac {\partial^{2}\phi}{\partial t^2}-\frac{1}{1+v(x)}\frac {\partial^{2}\phi}{\partial x^2}+\frac {\partial U(\phi)}{\partial \phi}=0.
\end{equation} 
The field energy density is
\begin{equation}\label{h2}
{\cal H}_{2}=\left(g^{00}(x)\right)^2 \frac{1}{2}\dot{\phi}^2+\frac{1}{2}\acute{\phi}^2+g^{00}(x)U(\phi).  
\end{equation}
The energy density is calculated by varying both the field and the metric (See page 643 equation (11.81) of \cite{i16}.). 
 
Solution (\ref{kg 1s}) can be used as initial condition for solving (\ref{new Equation motion}) when the potential $v(x)$ is small.
%''''''''''''''''''''''''''''''''''''''''''''''''''''''''''''''''''''''''''''''''''''''''''''''''''''''''''''''''''''''''''''''''''''''''''''''''''''''''''''''''''''''''''''''''''''''''''''''''''''''''''''''''''''''''''''''''''''''''''''''''''''''''''''''''''''''''''' 
\section {Numerical Simulations}
\quad\, Several simulations with different types of external potential $v(x)$ have been performed using two presented models. The smooth and slowly varying potential $v(x)=a e^{-b(x-c)^2}$ has been used in simulations which are reported below. Parameter "a" controls the strength of the potential, "b" represents its width and "c" adjusts the center of the potential. If $ a>0$, the potential shows a barrier and for $ a<0$ we have a potential well.

Simulations have been performed using 4th order Runge-Kutta method for time derivatives. Space derivatives were expanded using finite difference method. Grid spacing has been taken h=0.01, 0.02 and sometimes h=0.001. Time steps have been chosen as $\frac{1}{4}$ of the space step "h" because of numerical stability considerations. Simulations have been set up with fixed boundary conditions and solitons have been kept far from the boundaries. We have controlled the results of simulations by checking the total energy as a conserved quantity during the simulation.
 
A moving soliton has kinetic and potential energies. Consider a static (zero velocity) soliton, in this situation the soliton has only potential energy. The difference between the energies of the two static solitons in different positions comes from the difference between the potentials in those places.  This means that one can find the potential as a function of collective coordinate “X” using the energy of a static soliton in a different position “X”. The shape of the potential (as a function of X) has been found by placing a static soliton at different positions and calculating its energy. It is clear that its energy is equal to the potential in that position. Figure 2 shows the shape of the potential barrier $v(x)=0.5 e^{-4x^2}$ in different models which has been calculated using this method. This figure shows that the static parts of the potential is almost the same in both models. But we will find that the dynamical behaviour of the solitons are very different in the above models.

\begin{figure}[htbp]
\begin{center}
\leavevmode \epsfxsize=9cm \epsfbox {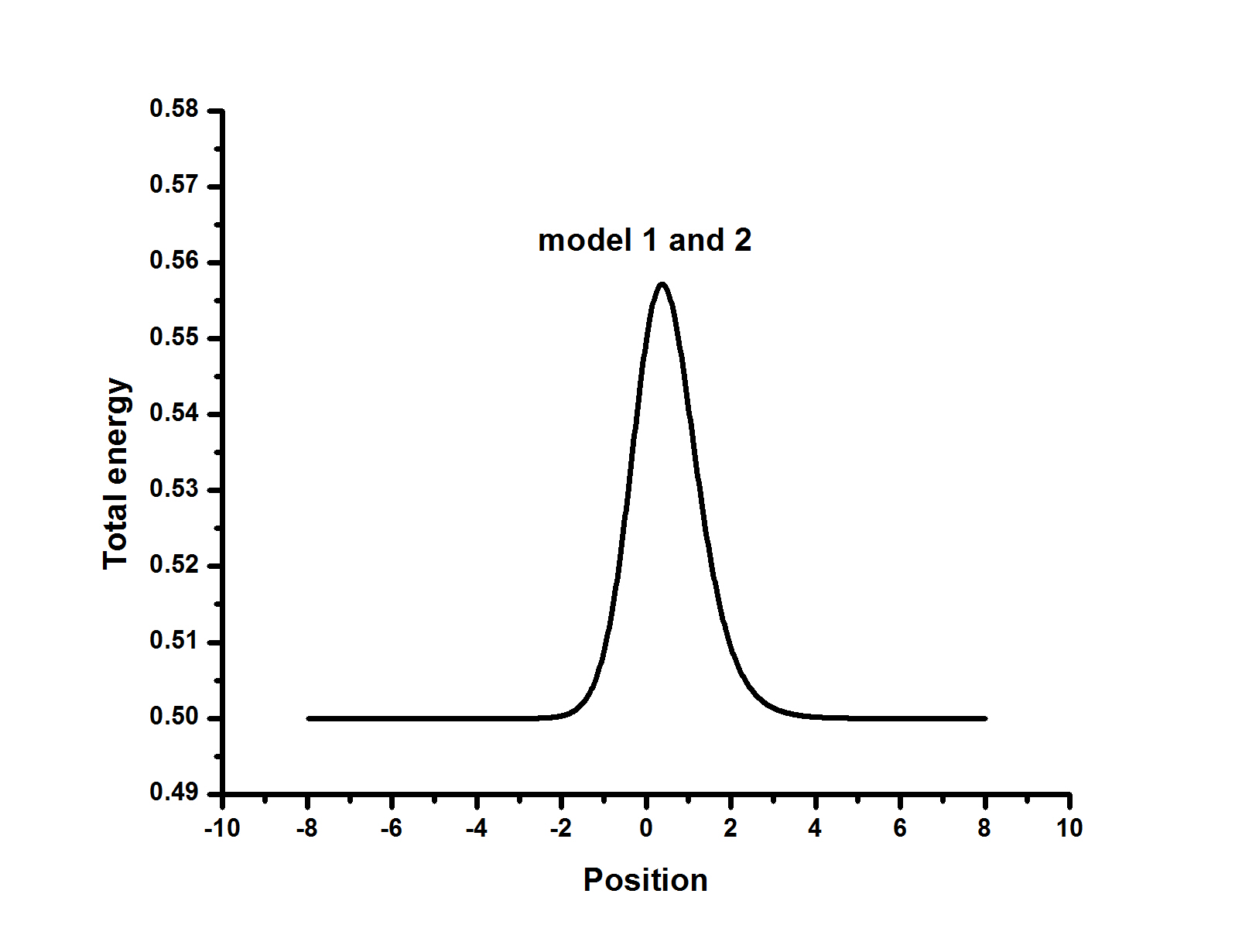}
\end {center}
\caption{Potential barrier $v(x)=0.5 e^{-4x^2}$as seen by the soliton in two models.}
\label{fig2}
\end{figure}

Suppose a soliton is placed far away from the potential. It goes toward the barrier in order to interact with the potential. There exist two different kinds of trajectories for the soliton during the interaction with the barrier depending on its initial velocity, which are separated by a critical velocity $u_{c}$. In low velocities $u_{i}<u_{c}$, soliton reflects back and reaches its initial place with final velocity of $u_{f}\approx u_{i}$. Figure 3 presents trajectories of a soliton with different values  of initial velocity for model 2. Critical velocity can be found by sending a soliton toward the potential with different initial velocities.

\begin{figure}[htbp]
\begin{center}
\leavevmode \epsfxsize=9cm \epsfbox {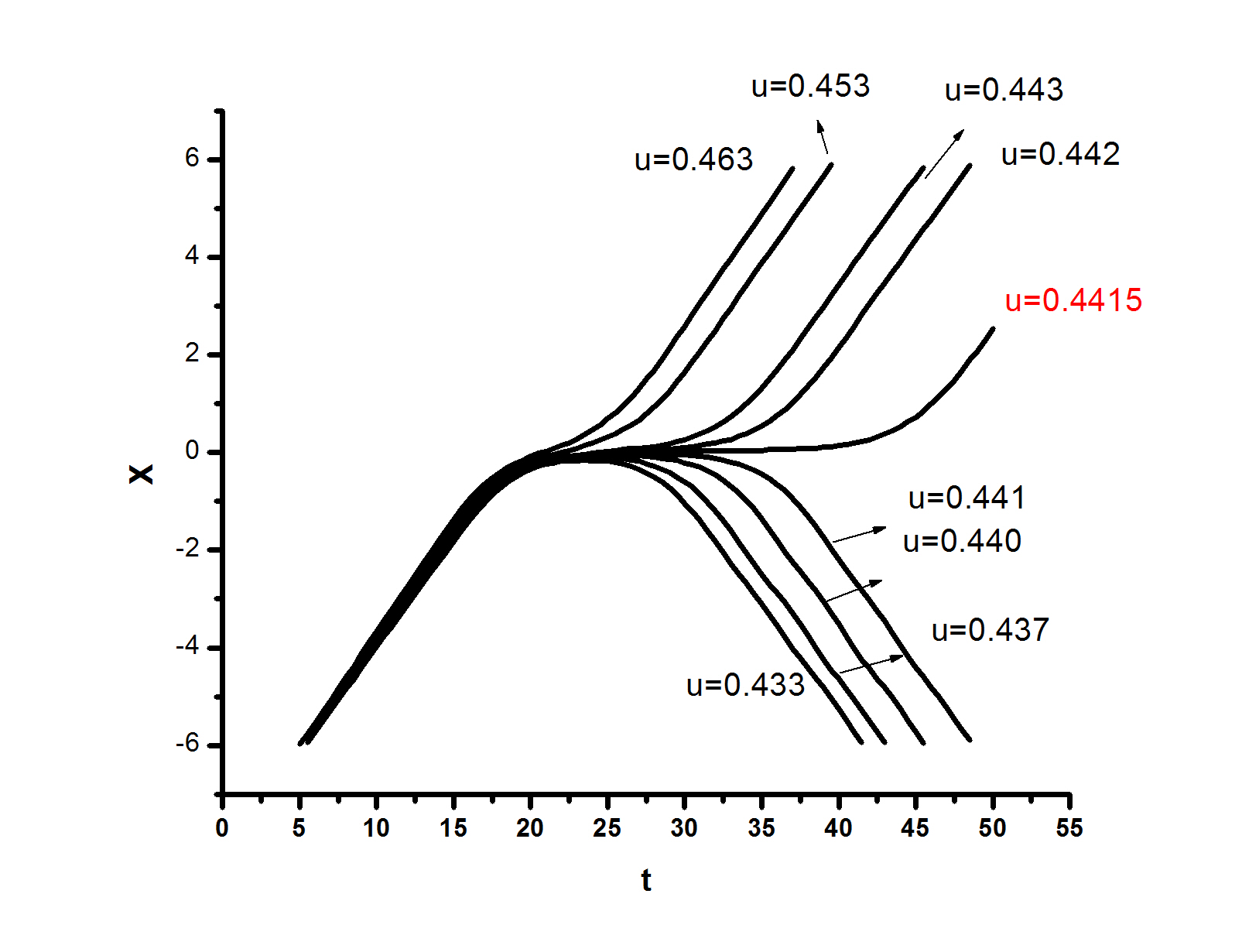}
\end {center}
\caption{Soliton trajectory during the interaction with the potential barrier $v(x)=0.5 e^{-4x^2}$ with different initial velocities for model 2.}
\label{fig3}
\end{figure}

Figure 4 presents the critical velocity as a function of the potential height in two models. This figure shows that the general behaviour of the soliton in the two models is the same. However, there is a little difference between the results of model 1 and 2. This originates from the difference between the rest  mass of the soliton in two models. Let us compare the soliton energy in model 1 (equation (\ref{h1})) and model 2 (equation (\ref{h2})). Static parts of the energy in two models are almost the same while the kinetic energy in two models is different. The effective mass of the soliton in model 1 is  $M_{eff1}=1$  while the effective mass in model 2 is $M_{eff2}=(1+v(x)) ^2$. The soliton rest mass of model 2 is greater than the soliton rest mass of model 1. The difference between the soliton rest mass in the two models increases when the potential height becomes greater.Thus the difference between the soliton critical velocity in two models become larger, which is clearly shown in figure 4.    

Simulations show that scattering of a soliton with a potential barrier is nearly elastic. The soliton radiates a small amount of energy during the interaction. The radiated energy during the interaction in models 1 and 2 is almost the same.

\begin{figure}[htbp]
\begin{center}
\leavevmode \epsfxsize=9cm \epsfbox {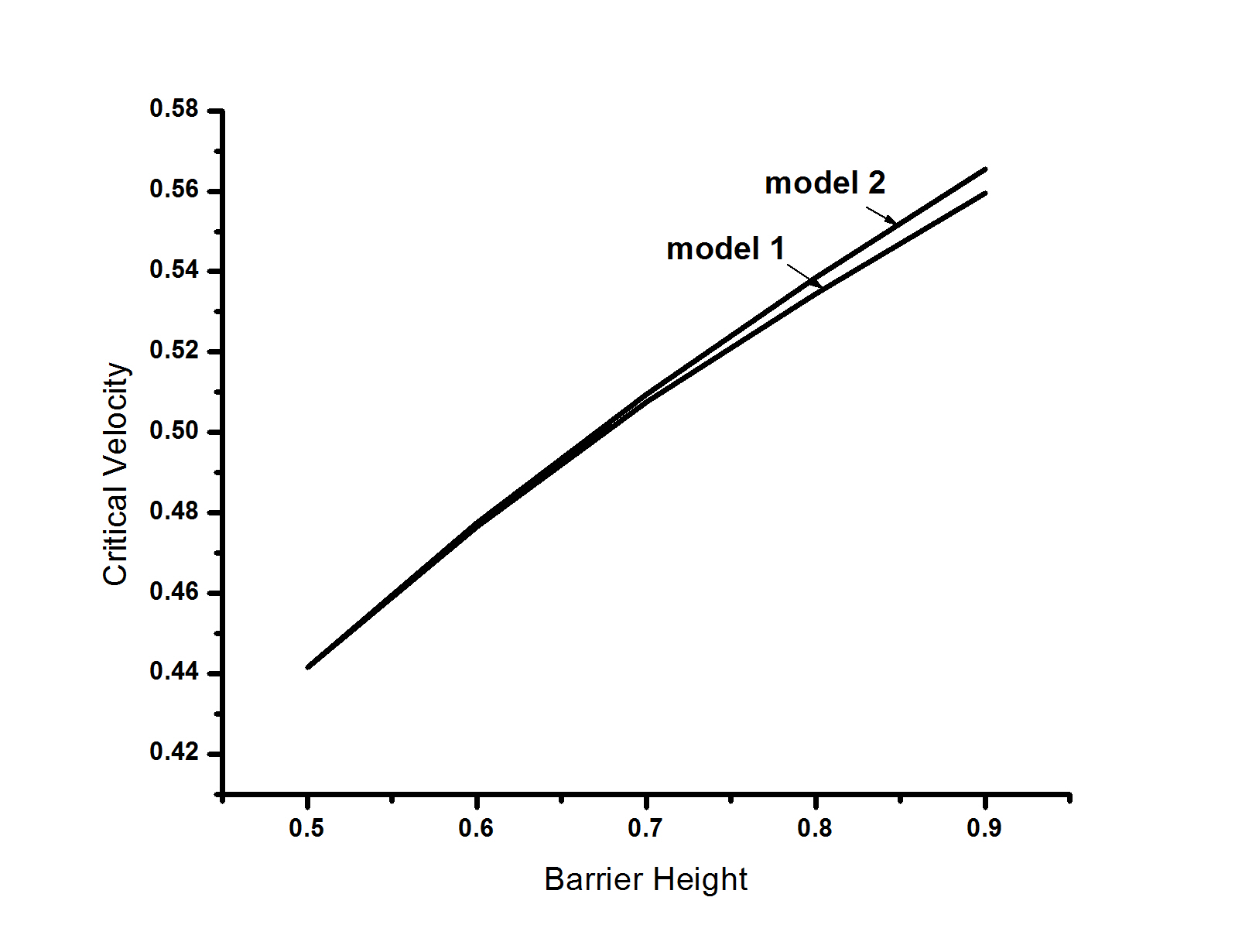}
\end {center}
\caption{Critical velocity as a function of the barrier height of the potential $v(x)=0.5 e^{-4x^2}$ in two models.}
\label{fig4}
\end{figure}

Scattering of topological solitons on a potential well is more interesting. Unlike a classical point particle which always passes through a potential well, a soliton may be trapped in a potential well with specific depth\cite{i17,i18}. It is shown that the $ \phi^4$ soliton has not a fixed mass during the interaction with a potential well \cite{i19} and it works for the NKG soliton too. So we cannot look at the soliton as a point particle in some cases. Several simulations have been done for the NKG soliton-well system using  the two models. Like the case of potential barrier, the general behaviour of the system is almost the same in both models. However, there are some differences in details of the interactions. Figure 5 presents a comparison between the shapes of the potential well $v(x)=-0.5e^{-4x^2}$ as seen by a  NKG soliton in two models. There is not a critical velocity for a soliton-well system, but we can define an escape velocity. Consider a soliton which is located in the initial position $x_{0}$ near the center of the potential $v(x)$. As figure 5 clearly shows, the soliton energy is a function of its initial position. This soliton can escape to infinity if its initial speed becomes greater than an escape velocity. The escape velocity for this situation is defined as the one for which the asymptotic speed is null at infinity. A soliton with initial velocity above the escape velocity, passes through the well and a soliton with initial velocity lower than the escape velocity falls into the well and is trapped by the potential.
\begin{figure}[htbp]
\begin{center}
\leavevmode \epsfxsize=9cm \epsfbox {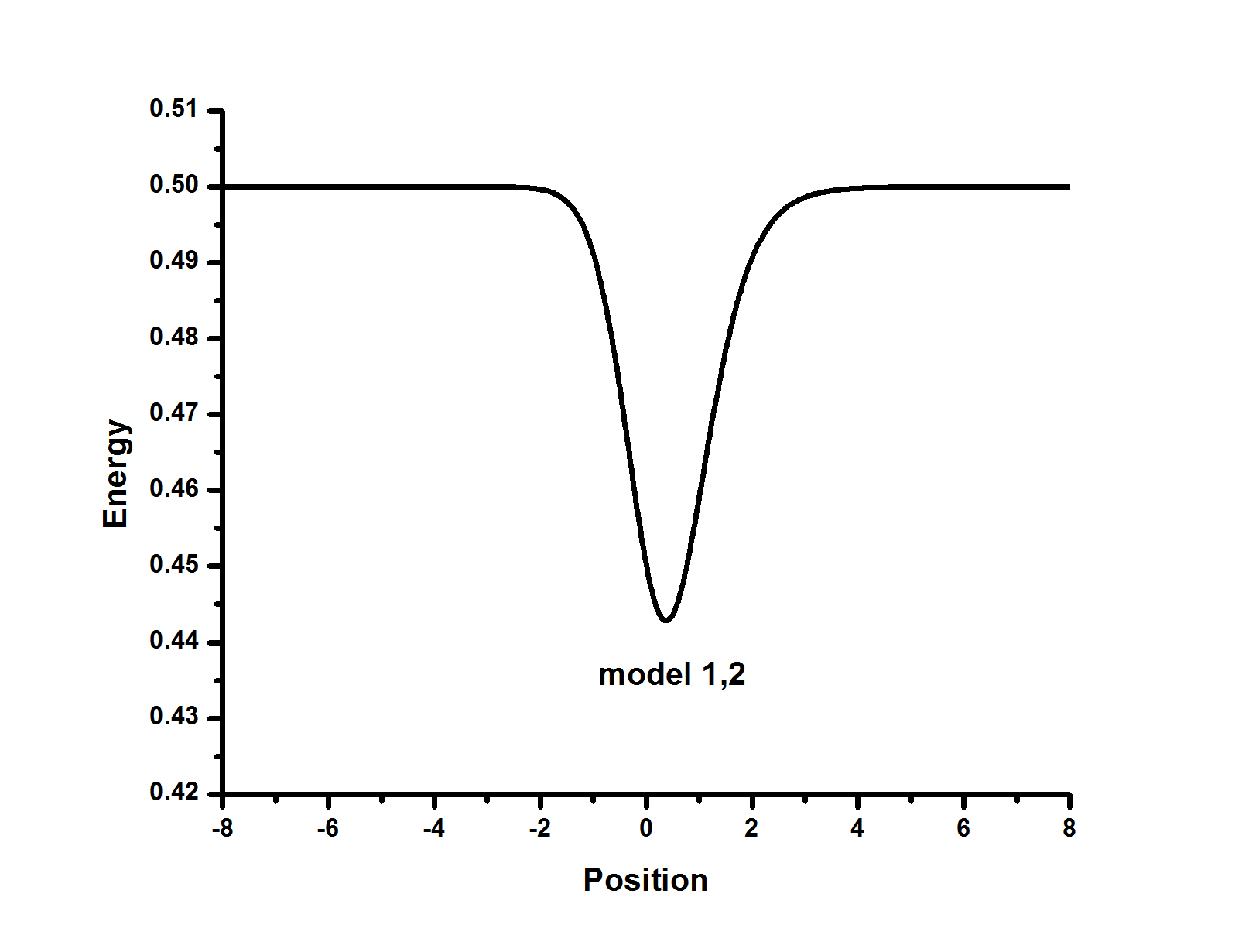}
\end {center}
\caption{The shape of the potential as seen by the soliton for the potential $v(x)=-0.5 e^{-4x^2}$}
\label{fig5}
\end{figure}

   The rest mass can be calculated with the integration of Homiltonian density (Eq.(\ref{h1}) for model 1, Eq.(\ref{h2}) for model 2) respect to the position $x$. The calculated rest mass using models 1 and 2 are very near to each other. The rest mass in two models are exactly the same because of the same static part of the energy in two models. Integration of Hamiltonian density is calculated by Rumberg method for different values of the potential height.

 As figure 5 shows the effective potential for these two models are the same. Therefore it is expected that the escape velocity of the soliton in models 1 and 2 finds the same values for different values of the potential height. Figure 6 presents the escape velocity of a soliton in the potential well as a function of the potential depth. However, the potentials in the models are the same, but the calculeted escape velocities using these models are not equal. The reason for this difference can be explained using figure 7 as follows.  

\begin{figure}[htbp]
\begin{center}
\leavevmode \epsfxsize=9cm \epsfbox {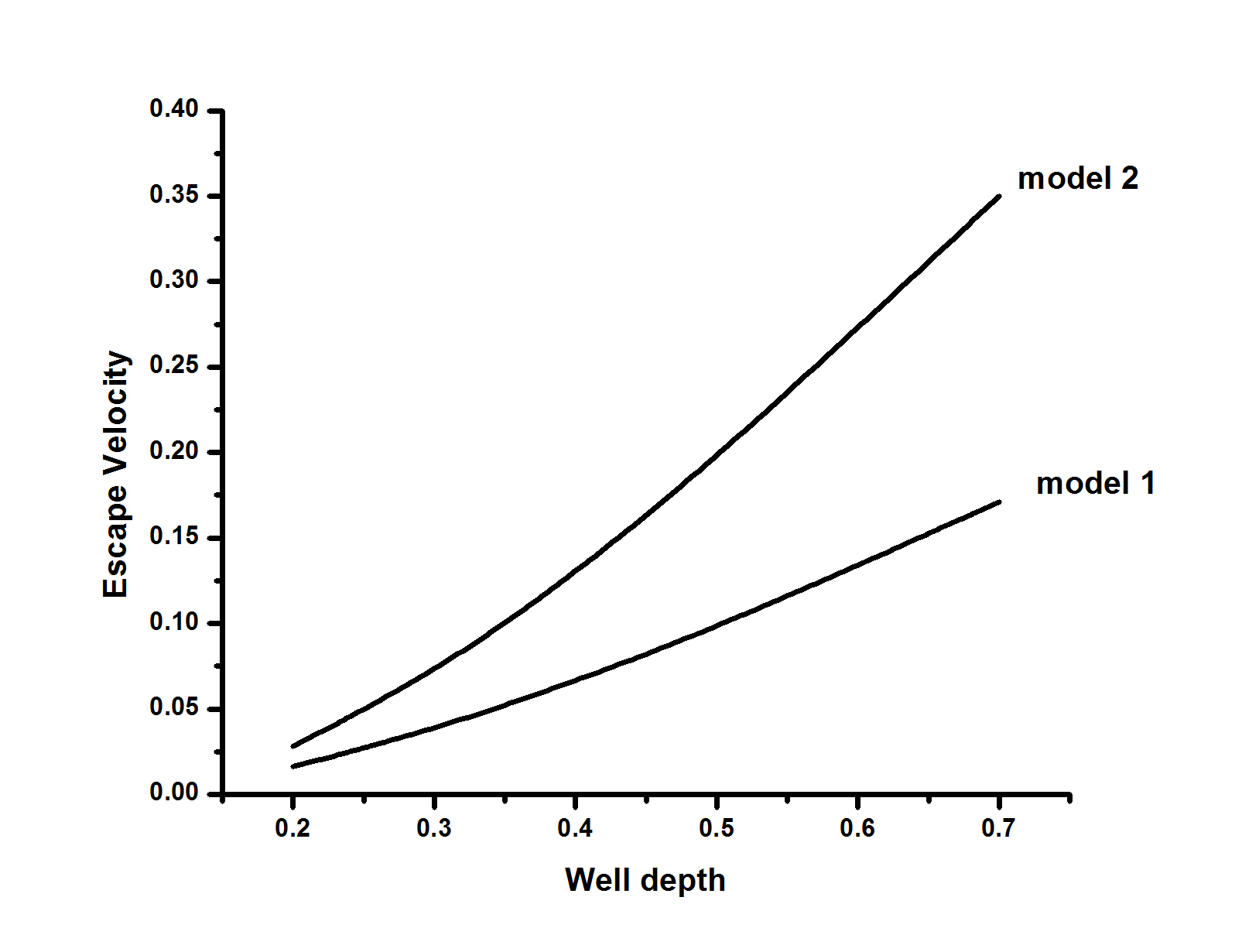}
\end {center}
\caption{Velocity as a function of the depth of the potential well.}
\label{fig7}
\end{figure}

Figure 7 presents the trajectory of a soliton with an initial velocity u=0.35 during the interaction with potential $v(x)=-0.5 e^{-4x^2}$. The final velocity of the soliton in model 2 is less than the final velocity of the soliton in model 1. This means that the energy loss due to radiation in model 2 is  greater than model 1. This is the main reason for the differences between the critical velocities observed in figure 6.

\begin{figure}[htbp]
  \begin{center}
   \leavevmode
 \epsfxsize=9cm   \epsfbox{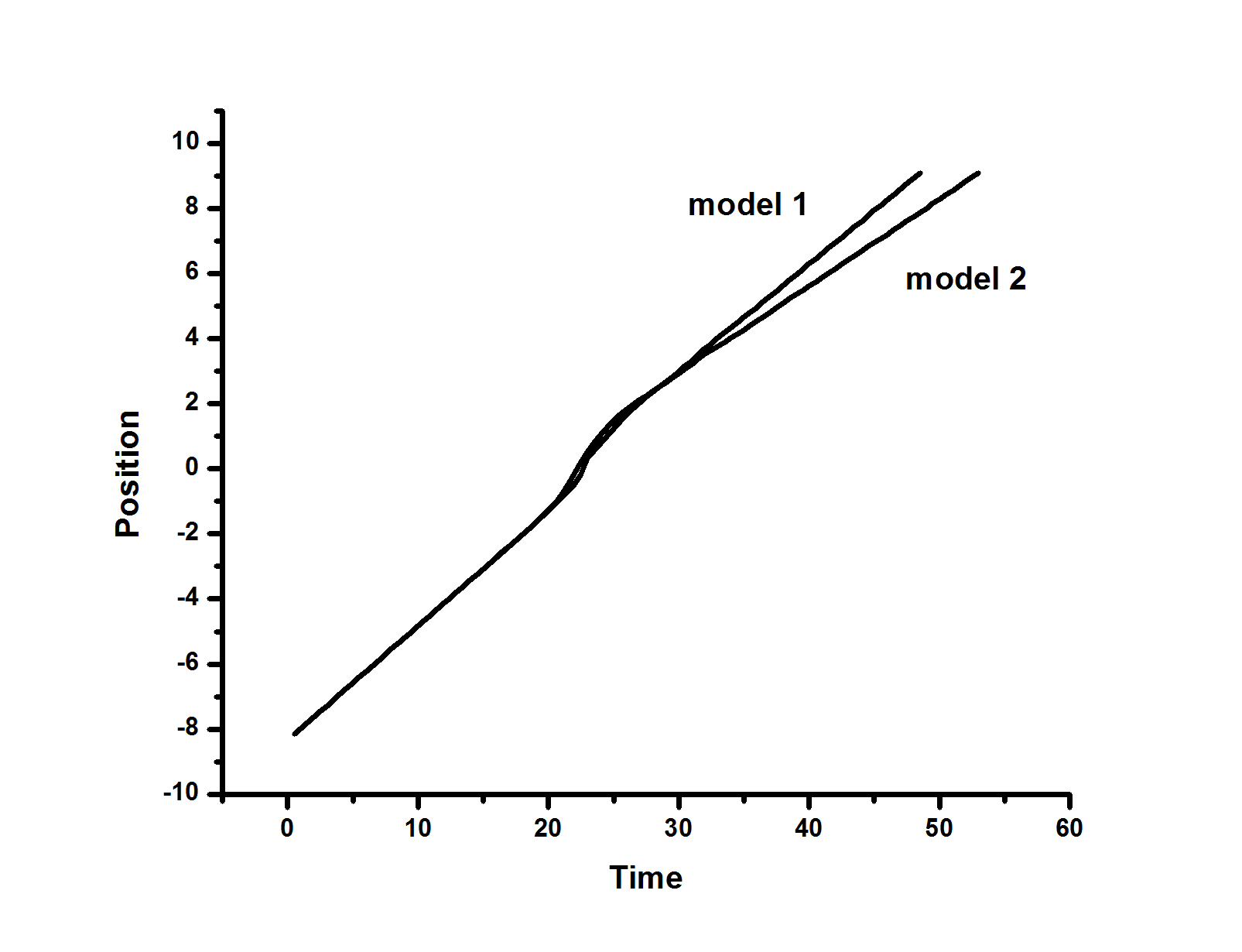}
  \end{center}
  \caption{Trajectory of a soliton with initial velocity u=0.35 during the interaction with the potential well $v(x)=-0.5 e^{-4x^2}$ in both models.  }
  \label{fig8}
\end{figure}

The most interesting behaviour of a soliton during the process of scattering on a potential well is seen in some very narrow windows of initial velocities. At some velocities less than the $u_c$ the soliton may reflect back or pass over the potential while one would expect that the soliton should be trapped in the potential well. These narrow windows can be found by scanning the soliton initial velocity with small steps. Figures 9 presents this phenomenon in models 1 and 2. Figure 8(a) shows that a soliton with an initial velocity within the window  $0.062\leq u_{i}\leq 0.065$ reflects back after the interaction with the potential $v(x)=-0.4 e^{-4x^2}$ simulated using model 1. Figure 8(b) presents the same situation simulated using model 2. We couldn't find a soliton reflection in model 2 for the potential $v(x)=-0.4 e^{-4x^2}$. Some simulations have been performed for different values of the potential height (a=-0.2,-0.3,-0.5,-0.6,-0.7,-0.8) in model 2, but such a soliton reflection from the potential well has not been observed at all. Model 2  is built by varying the Lagrangian density with respect to both "field" and "metric", therefore an energy exchange between the field and the potential is possible in this model. Some authors have related such a soliton reflection to the energy exchange between the field and the potential \cite {i19, i20}.   It seems that this point is not true for the NKG solitons. Our simulations for the NKG solitons did not confirm this situation for model 2. This phenomenon needs more investigations. 

\begin{figure}[htbp]
  \begin{center}
 \leavevmode
 \epsfxsize=7cm   \epsfbox{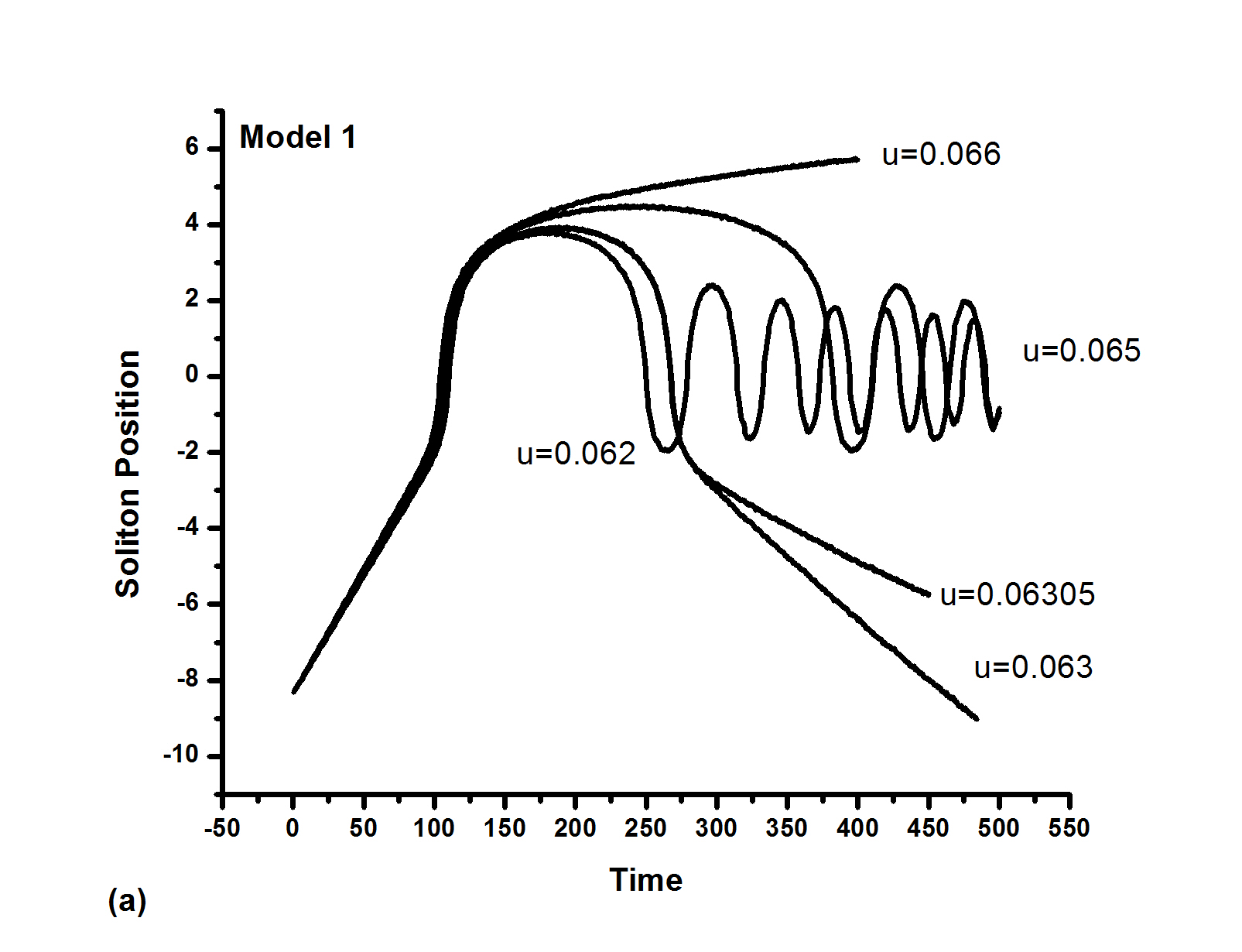}\epsfxsize=7cm \epsfbox{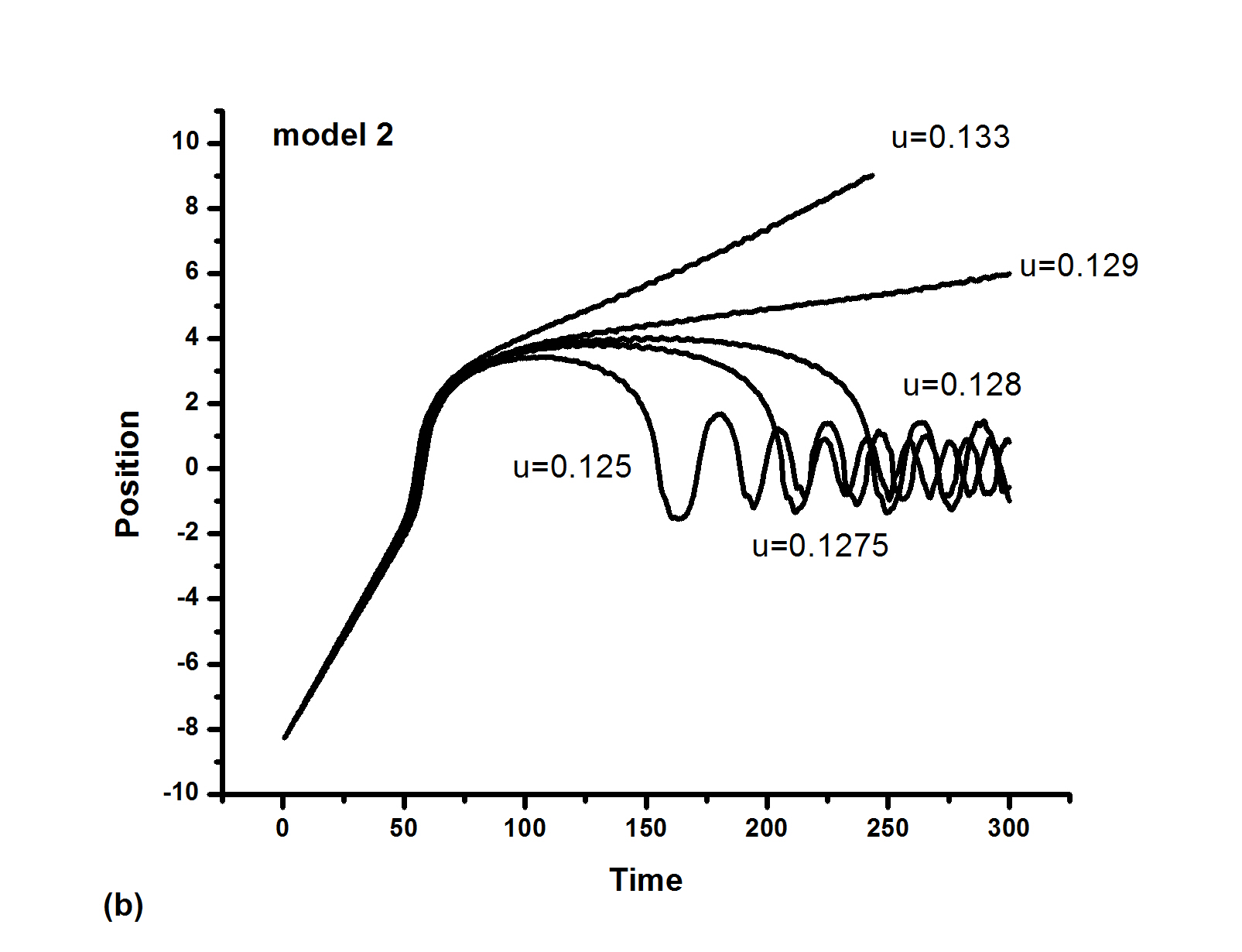} 
  \end{center}
  \caption{(a) Soliton reflection from the potential well $v(x)=-0.4 e^{-4x^2}$ in model 1. (b) Soliton reflection from the potential well $v(x)=-0.4 e^{-4x^2}$ in model 2.  }
  \label{fig:fig9}
\end{figure}

%====================================================================

\section{Conclusion}

Two models have been used to study the interaction of the NKG solitons with defects. The general behaviour of a soliton during the interaction with external potentials in two models are almost the same. Model 1 adds the potential to the equation of motion by a different method from the one that is used in the model 2. Model 1 is easy to simulation while model 2 is more analytic.The two models confirm that the interaction of a soliton with a potential barrier is nearly elastic. At low velocities it reflects back but with a high velocity climbs the barrier and passes over the potential. Soliton radiates some amounts of energy during the interaction with the potential. There exists a critical velocity which separates these two kinds of trajectories. The interaction of a soliton with the potential well is more inelastic. It is possible that a soliton scatters on a potential well and reflects back from the potential. This phenomenon depends on the model. Model 1  predicts this behaviour, but this is in contradiction with what can be seen for the Sine-Gordon and $\phi^4 $ models. These models show this behaviour for model 2 while such a behaviour is observed in model 1 for the NKG solitons.

It is interesting to investigate scattering of solitons of other models on defects using model 2. These studies help us to gain more comprehensive knowledge about the general behaviour of solitons.
%====================================================================

\end{document}